\documentclass{article}

\usepackage{amsmath,amssymb,amsfonts}
\usepackage{a4wide}
\usepackage{color}
\usepackage{bm}
\usepackage{graphicx}
\usepackage{euscript}
\usepackage{ulem}
\usepackage[version=3]{mhchem}

\definecolor{gray}{gray}{0.6}

\date{\today}
\title{Exact steady states of the impurity-doped $XXZ$ spin chain coupled to dissipators}
\author{Chihiro Matsui$^1$ and Naoto Tsuji$^{2,3}$ \\[3ex]
{\it $^1$Graduate School of Mathematical Sciences, The University of Tokyo}\\
{\it 3-8-1, Komaba, Meguro-ku, 153-8914 Tokyo, Japan}\\
{\it $^2$Department of Physics, The University of Tokyo} \\
{\it 7-3-1, Hongo, Bunkyo-ku, 113-0033 Tokyo, Japan}\\
{\it $^3$RIKEN Center for Emergent Matter Science (CEMS)}\\
{\it 2-1, Hirosawa, Wako, 351-0198 Saitama, Japan}
}

\begin{document}
\maketitle

\begin{center}
{\bf Abstract}
\end{center}
\bigskip
{\small
We give an exact matrix product steady state and matrix product forms of local observables for the bulk impurity-doped $XXZ$ spin model coupled to dissipators at both ends, whose dynamics is described by the Lindblad quantum master equation. We find that local magnetization is induced at the impurity site when the spin current flows, which is contrary to the usual situation where current suppresses magnetization due to heating. It is proved that this current-induced magnetization survives in the thermodynamic limit, and the spin current does not depend on the impurity strength. We discuss the role of bulk integrability by comparing the results with those of non-integrable models solved numerically by the quantum trajectory method.
}

\section{Introduction}
Transport is one of the most fundamental properties of nonequilibrium quantum many-body systems. It reflects thermalization behavior and integrability of those systems \cite{bib:KWW06, bib:RDYO07}. 
For example, macroscopic current approaches zero in the long-time limit in generic isolated non-integrable systems, while transport becomes ballistic and current may not vanish in the long-time steady state in integrable isolated systems
\cite{bib:BBNVBB76, bib:CZP95, bib:FK03, bib:GM92, bib:JR07, bib:M20, bib:PBLBT06, bib:P11, bib:PI13, bib:RA00, bib:S06, bib:SPA09, bib:THIC01, bib:Z99, bib:ZNP97}. 
For open quantum systems, which is our focus in the paper, steady state current is observed both in bulk integrable and non-integrable systems. However, there could be a fundamental difference between them, which is still to be understood. An interesting question in this context is how an impurity inserted in a {\it bulk integrable} quantum system can change the behavior of current and other observables {\it far from equilibrium with dissipation}. 

The integrability has played an important role in understanding nonequilibrium open quantum systems. When the system deviates far from equilibrium, it is often difficult to calculate observables exactly in nonequilibrium steady states in the thermodynamic limit. In recent years, on the other hand, there have been various studies on exactly solvable steady states of dissipative quantum systems. Especially for systems whose time evolution is described by the Lindblad master equation, several mechanisms have been discussed on the solvability of steady states: In some cases, the full spectrum of the Liouvillian is accessible via free fermionization \cite{bib:P08, bib:PZ10, bib:SK19, bib:V20}, triangularization of the Liouvillian operator \cite{bib:BBMJ20, bib:NKU21}, operator space fragmentation \cite{bib:EP20}, and the Bethe ansatz method \cite{bib:MEP16, bib:SK19, bib:ZE19}. Other cases allow us to know only the steady state in the matrix product form due to bulk integrability under the proper choice of boundary dissipators \cite{bib:BD15, bib:I17, bib:IP14, bib:IZ14, bib:KPS13, bib:MP17, bib:PP15, bib:PKS13, bib:PPZ20-2, bib:PPZ20, bib:PZP22, bib:PKS19, bib:P11-2, bib:P11, bib:P14, bib:P14-2, bib:P15, bib:PIP13, bib:PZ09}. In this paper, we focus on the latter case. Initially, the matrix product form of the steady-state solution has heuristically been found for the Lindblad master equation on the $XXZ$ spin chain with dissipators attached at both ends \cite{bib:P11-2, bib:P11}. Later, each matrix element in the matrix product state has been found to be related to the Lax operator of the bulk Hamiltonian \cite{bib:IZ14, bib:PIP13}. This remarkable fact opened the way to study steady states in a wide class of dissipative quantum systems with bulk integrability.

To better understand the role of integrability on transport properties, we study the effect of impurities in integrable and non-integrable open quantum systems. 
Previously, transport properties of many-body systems with an impurity have been studied, e.g., for a one-dimensional Luttinger liquid \cite{bib:KF92b, bib:KF92}, the Anderson impurity model \cite{bib:HDW91, bib:MW92, bib:MWL93, bib:NL88}, and a cold-atom gas with an impurity atom \cite{bib:SKIYCD18}.
In this paper, we focus on the $XXZ$ spin chain coupled to dissipators at both ends, whose steady state is written in the matrix product form \cite{bib:IZ14, bib:P11-2, bib:P11, bib:PIP13}. It is known that the $XXZ$ model can be deformed to include impurities without breaking its integrability \cite{bib:FZ97, bib:S94, bib:ZS97},
where the leading coupling to an impurity is represented by the (anisotropic) triple spin scalar product, which is in sharp contrast to other impurity models.
For integrability-preserving models, we find that the exact steady state is still given in a matrix product form of the Lax operators. Based on this, we calculate the current and magnetization exactly, which can be evaluated in the thermodynamic limit. The result shows that local magnetization is generated at the impurity site when the current flows through the system. This behavior is rather opposite to our naive expectation that current generally suppresses magnetization via heating. We also study the difference between integrable and non-integrable systems by numerical simulations of finite-size systems via the quantum trajectory method \cite{bib:Carmichael93, bib:Daley14, big:DCM92}.

This paper is organized as follows. In the next section, we define the dissipative $XXZ$ model with bulk impurity. At the same time, we also give its properties from integrability aspect. In section 3, we construct the steady state in the matrix product form. Basic physical quantities are analytically calculated on the steady state, whose results are supported by numerical calculations. The last section is devoted to the concluding remarks and future problems. 

\section{Impurity-doped $XXZ$ spin chain}

\subsection{Quantum spin system coupled to dissipators}

We consider the $XXZ$ spin chain coupled to dissipators at both ends. The dissipators at the boundaries serve as ``spin baths'', which provides up spins from the left boundary and down spins from the right boundary, respectively. In the bulk chain, we insert an impurity centered around a certain site. Let us denote the bulk Hamiltonian including the impurity by $H_{XXZ}^{\rm imp}$, the detail of which is described in the next section. We assume that the time evolution of the density matrix of the system, $\rho$, is given by the following Lindblad master equation,
\begin{align}
	&\frac{d}{dt} \rho(t) = \mathcal{L}(\rho(t)), \label{eq:LB}\\
	&\mathcal{L}(\rho) = -i [H_{XXZ}^{\rm imp},\,\rho] + \sum_{\mu \in \{{\rm L,R}\}} \varepsilon_{\mu} \mathcal{D}_{\mu}(\rho). 
    \label{eq:LB2}
\end{align}
Here $\mathcal L$ is the Liouvillian superoperator,
$\varepsilon_\mu$ ($\mu = {\rm R,L}$) is the dissipation rate,
and $\mathcal{D}_{\mu}$ ($\mu = {\rm R,L}$) is the dissipation operator which maps the density matrix as 
\begin{equation}
	\mathcal{D}_{\mu}(\rho) = 2A_{\mu} \rho A_{\mu}^{\dag} - \{A_{\mu}^{\dag} A_{\mu},\, \rho\}, 
\end{equation}
where we take $A_{\rm L} = \sigma_1^+$ and $A_{\rm R} = \sigma_N^-$ representing the effect of spin injection at the boundaries. In this setup, the dynamics satisfies Markovianity, i.e., $\rho(t+dt)$ is determined solely from $\rho(t)$ and does not depend on the previous history of the evolution. The justification of the Markovian property often requires several physical assumptions, such as the weak system-bath coupling and short correlation time of the baths, which, however, we do not discuss in detail here.
In the following discussion, we adopt the Lindblad equation (\ref{eq:LB}) and (\ref{eq:LB2}) that are supposed to be valid for arbitrary $\varepsilon_\mu$, and assume that the impurity is not located at the ends of the chain.

\subsection{The $XXZ$ spin chain}
The anisotropic Heisenberg spin chain, the so-called $XXZ$ spin chain, with the open boundary condition is given by the following Hamiltonian,
\begin{align}
	&H_{XXZ} = \sum_{j=1}^{N-1} h_{j,j+1}(\Delta), \\
	&h_{j,j+1}(\Delta) = \sigma_j^x \sigma_{j+1}^x + \sigma_j^y \sigma_{j+1}^y + \Delta \sigma_j^z \sigma_{j+1}^z.
\end{align} 
The Hamiltonian $H_{XXZ}$ acts on the Hilbert space $\mathfrak{H}$ spanned by $\otimes_{j=1}^N \mathfrak{h}_j$ with $\mathfrak{h}_j = \mathbb{C}^2$, where the operators $\sigma_j^{\alpha}$ ($\alpha = x,y,z$) nontrivially act on the $j$th space,
\begin{equation}
	\sigma_j^{\alpha} = \bm{1} \otimes \dots \otimes \bm{1} \otimes \underset{j\text{th}}{\sigma^{\alpha}} \otimes \bm{1} \otimes \cdots \otimes \bm{1}. 
\end{equation}
The operators $\sigma^{\alpha}:\mathbb{C}^2 \to \mathbb{C}^2$ are the Pauli matrices satisfying the $\mathfrak{sl}_2$ relations:
\begin{equation}
    [\sigma^{\alpha},\,\sigma^{\beta}] = 2i \varepsilon_{\alpha\beta\gamma} \sigma^{\gamma},
\end{equation}
where $\varepsilon_{\alpha\beta\gamma}$ is the Levi-Civita symbol. 
The $XXZ$ spin chain is well-known to be integrable. This model shows gapless excitations for the anisotropy $\Delta \leq 1$ and gapped excitations for $\Delta > 1$ in the thermodynamic limit $N \to \infty$. In the following, 
we use the conventional parametrization $\Delta = \cosh\eta$. $\eta$ takes a real value in the gapped regime, while it takes a pure imaginary value in the gapless regime. In the gapless regime, we sometimes denote $\eta = i\gamma$ for simplicity. 

By doping the impurity at the $n$th site, the $XXZ$ Hamiltonian is deformed due to the interaction with the impurity,
\begin{equation}
	H_{XXZ}^{\rm imp} = \sum_{j=1 \atop j \neq n-1,n}^{N-1} \Big(h_{j,j+1}(\Delta) + h_{n-1,n,n+1}(\Delta,\xi)\Big).
 \label{eq: H imp}
\end{equation}
In this paper, we focus on two kinds of integrable interactions, $h_{n-1,n,n+1}(\Delta,\xi)=h_{n-1,n,n+1}^{\text{int}(1)}$ and $h_{n-1,n,n+1}(\Delta,\xi)=h_{n-1,n,n+1}^{\text{int}(2)}$. The first one of the integrable deformation is {\it the spin-$1/2$ impurity}: 
\begin{align} \label{eq:int-1a}
	h^{\text{int}(1)}_{n-1,n,n+1}
    &= 
    \frac{1 - \Delta^2}{1 - \Delta^2 + \sinh^2\xi}
    \Big(\cosh\xi(\sigma_{n-1}^x \sigma_n^x + \sigma_{n-1}^y \sigma_n^y) + \Delta \sigma_{n-1}^z \sigma_n^z \Big) \nonumber
    \\
    &\quad
    +\frac{1 - \Delta^2}{1 - \Delta^2 + \sinh^2\xi}
    \Big(\cosh\xi (\sigma_n^x \sigma_{n+1}^x + \sigma_n^y \sigma_{n+1}^y) + \Delta \sigma_n^z \sigma_{n+1}^z \Big) \nonumber
    \\
    &\quad
    +\frac{1}{1 - \Delta^2 + \sinh^2\xi}
    \Big( \Delta \sinh^2\xi (\sigma_{n-1}^x \sigma_{n+1}^x + \sigma_{n-1}^y \sigma_{n+1}^y + \sigma_{n-1}^z \sigma_{n+1}^z ) \nonumber
    \\
    &\hspace{31mm}
    +\Delta\sqrt{1 - \Delta^2} \sinh\xi (\sigma_{n-1}^x \sigma_n^y - \sigma_{n-1}^y \sigma_n^x) \sigma_{n+1}^z \nonumber
    \\
    &\hspace{31mm}
    +\Delta\sqrt{1-\Delta^2} \sinh\xi
    \cdot \sigma_{n-1}^z (\sigma_n^x \sigma_{n+1}^y - \sigma_n^y \sigma_{n+1}^x) \nonumber
    \\
    &\hspace{31mm}
    -\sqrt{1-\Delta^2} \sinh\xi \cosh\xi
    (\sigma_{n-1}^x \sigma_n^z \sigma_{n+1}^y - \sigma_{n-1}^y \sigma_n^z \sigma_{n+1}^x)
    \Big)
\end{align}
for $\Delta\le 1$, and
\begin{align} \label{eq:int-1b}
    h^{\text{int}(1)}_{n-1,n,n+1}
    &=
    \frac{\Delta^2 - 1}{\Delta^2 -1 + \sin^2\xi}
    \Big(\cos\xi(\sigma_{n-1}^x \sigma_n^x + \sigma_{n-1}^y \sigma_n^y) + \Delta \sigma_{n-1}^z \sigma_n^z \Big) \nonumber
    \\
    &\quad
    +\frac{\Delta^2 - 1}{\Delta^2 -1 + \sin^2\xi}
    \Big(\cos\xi (\sigma_n^x \sigma_{n+1}^x + \sigma_n^y \sigma_{n+1}^y) + \Delta \sigma_n^z \sigma_{n+1}^z \Big) \nonumber
    \\
    &\quad
    +\frac{1}{\Delta^2 -1 + \sin^2\xi}
    \Big( \Delta \sin^2\xi (\sigma_{n-1}^x \sigma_{n+1}^x + \sigma_{n-1}^y \sigma_{n+1}^y + \sigma_{n-1}^z \sigma_{n+1}^z ) \nonumber
    \\
    &\hspace{31mm}
    +\Delta\sqrt{\Delta^2 - 1} \sin\xi (\sigma_{n-1}^x \sigma_n^y - \sigma_{n-1}^y \sigma_n^x) \sigma_{n+1}^z \nonumber
    \\
    &\hspace{31mm}
    +\Delta\sqrt{\Delta^2 - 1} \sin\xi
    \cdot \sigma_{n-1}^z (\sigma_n^x \sigma_{n+1}^y - \sigma_n^y \sigma_{n+1}^x) \nonumber
    \\
    &\hspace{31mm}
    -\sqrt{\Delta^2 - 1} \sin\xi \cos\xi
    (\sigma_{n-1}^x \sigma_n^z \sigma_{n+1}^y - \sigma_{n-1}^y \sigma_n^z \sigma_{n+1}^x)
    \Big)
\end{align}
for $\Delta>1$.
The parameter $\xi$ is the so-called {\it inhomogeneous parameter}, which is introduced to represent the effect of the impurity (i.e., the impurity strength). In order for the Hamiltonian to be Hermitian, we choose $\xi$ as a real value. The Hamiltonian for $\Delta \leq 1$ can be mapped to the one for $\Delta>1$ by replacing $\xi$ with $i\xi$. 
The isotropic limit ($\Delta\to 1$) becomes singular around $\xi=0$. However, if one takes the scaling $\xi = \gamma \xi'$ with $\xi'$ fixed, the isotropic limit of the impurity-doped model at $\Delta=\cos\gamma \to 1$ is well-defined. 

Although the Hamiltonians (\ref{eq:int-1a}) and (\ref{eq:int-1b}) seem to contain complicated terms, the form of the expression becomes much simplified for small $\xi$.
In fact, if one expands $h_{n-1,n,n+1}^{\rm int(1)}$ in terms of $\xi$, the leading terms read
\begin{align}
    h_{n-1,n,n+1}^{\rm int(1)}
    &= h_{n-1,n}(\Delta)+h_{n,n+1}(\Delta)+\frac{\xi}{2}\frac{1}{\sqrt{|1 - \Delta^2|}} (\bm{\sigma}_{n-1},\bm{\sigma}_n,\bm{\sigma}_{n+1})_{\Delta} + O(\xi^2),
\end{align}
where we have defined the {\it $\Delta$-deformed triple product},
\begin{equation}
	(\bm{A},\bm{B},\bm{C})_{\Delta} = 
	\Delta B^x (C^yA^z - A^yC^z)
	+ \Delta B^y (C^zA^x - A^zC^x)
	+ B^z (A^xC^y - C^xA^y),  
\end{equation}
for the three-component operators $\bm{A},\bm{B},\bm{C}$. It is an anisotropic generalization of the isotropic scalar triple product, $(\bm A, \bm B, \bm C)_{\Delta=1}=\bm A\cdot(\bm B\times\bm C)$.

The second integrable deformation is {\it the higher-spin impurity}: 
\begin{align} \label{eq:int-2}
h^{\text{int}(2)}_{n-1,n,n+1} 
&= \frac{1}{\sinh^2(\tfrac{3\eta}{2})}
    \Big\{ 
    \sinh^2\eta \cdot \sigma_{n-1}^+ \cosh(\eta (\tfrac{1}{2} + S_n^z)) S_n^-
    + \sinh\eta \cosh\eta \cdot \sigma_{n-1}^+ \sinh(\eta (\tfrac{1}{2} + S_n^z)) \cdot S_n^- \sigma_{n+1}^z
    \nonumber \\
    &+ \sinh^2\eta \cdot \sigma_{n-1}^- \cosh(\eta (\tfrac{1}{2} - S_n^z)) \cdot S_n^+
    - \sinh\eta \cosh\eta \cdot \sigma_{n-1}^- \sinh(\eta (\tfrac{1}{2} - S_n^z)) \cdot S_n^+ \sigma_{n+1}^z
    \nonumber \\
    &+ \sinh^2\eta \cosh(\eta (\tfrac{1}{2} + S_n^z)) \cdot S_n^- \sigma_{n+1}^+ 
    + \sinh\eta \cosh\eta \cdot \sigma_{n-1}^z \sinh(\eta (\tfrac{1}{2} + S_n^z)) \cdot S_n^- \sigma_{n+1}^+
    \nonumber \\
    &+ \sinh^2\eta \cosh(\eta (-\tfrac{1}{2} + S_n^z)) \cdot S_n^+ \sigma_{n+1}^-
    + \sinh\eta \cosh\eta \cdot \sigma_{n-1}^z \sinh(\eta (-\tfrac{1}{2} + S_n^z)) \cdot S_n^+ \sigma_{n+1}^-
    \nonumber \\
    &+ \frac{3}{4} \sinh\eta \cosh\eta \cdot \Big( \sigma_{n-1}^z \sinh(2\eta S_n^z) + \sinh(2\eta S_n^z) \cdot \sigma_{n-1}^z \Big)
    \nonumber \\
    &- \Big( \frac{1}{4} \cosh\eta \cdot (\cosh\eta \cosh(2\eta S_n^z) - 1) - \frac{1}{2} \sinh^2\eta \cosh(2\eta S_n^z) + \frac{1}{4} \sinh^2\eta \cosh\eta \cdot \{S_n^+,\, S_n^-\} \Big)
    \nonumber \\
    &- \sigma_{n-1}^z \Big( \frac{1}{4} \cosh\eta \cdot (\cosh\eta \cosh(2\eta S_n^z) - 1) - \frac{1}{4} \sinh^2\eta \cosh\eta \cdot \{S_n^+,\, S_n^-\} \Big) \sigma_{n+1}^z
    \nonumber \\
    &+ (\sinh\eta)^2 \sigma_{n-1}^+ (S_n^-)^2 \sigma_{n+1}^+
    + (\sinh\eta)^2 \sigma_{n-1}^- (S_n^+)^2 \sigma_{n+1}^-
    \Big\}. 
\end{align}
The spin operators $S_j^{\alpha}$ ($\alpha=+,-,z$), which non-trivially act on the $j$th space 
\begin{equation}
    S_j^{\alpha} = \bm{1} \otimes \cdots \otimes \bm{1} \otimes \underset{j\text{th}}{S^{\alpha}} \otimes \bm{1} \otimes \cdots \otimes \bm{1},  
\end{equation}
are the $U_q(\mathfrak{sl}_2)$ generators of an arbitrary $(2s+1)$-dimensional irreducible representations ($2s \in \mathbb{Z}_{>0}$) with the highest weight $s$: 
\begin{align} \label{eq:Uq-integer}
	&e^{\pm \eta S_a^z} = \sum_{m=0}^{2s} e^{\pm \eta (s-m)} |m \rangle \langle m|, \\
	&S_a^+ = \sum_{m=0}^{2s-1} \frac{\sinh(\eta(m+1))}{\sinh\eta} |m \rangle \langle m+1|, \quad
	S_a^- = \sum_{m=0}^{2s-1} \frac{\sinh(\eta(2s-m))}{\sinh\eta} |m+1 \rangle \langle m|. 
\end{align}
%
%

The first model among the integrable cases is {\it the spin-$1/2$ impurity}, around which the three spin-$1/2$ spins interact each other due to inhomogeneous impurity. The second one is {\it the higher-spin impurity}, which has an arbitrary spin interacting with the neighboring spin-$1/2$ spins. 


Besides the integrable deformation of the $XXZ$ model, we also consider two kinds of non-integrable interactions: One is the impurity interaction with the isotropic scalar triple product (model I),
\begin{align}
    &h^{\text{nint}({\rm I})}_{n-1,n,n+1} = h_{n-1,n}(\Delta) + \frac{\xi}{2}\frac{1}{\sqrt{|1-\Delta^2|}}
    (\bm\sigma_{n-1}, \bm\sigma_n, \bm\sigma_{n+1})_{\Delta=1} + h_{n,n+1}(\Delta), 
    \label{eq: nint I}
\end{align}
and the other is the interaction with a local magnetic field (model II),
\begin{align}
	&h^{\text{nint}({\rm II})}_{n-1,n,n+1} = h_{n-1,n}(\Delta) + \frac{\xi}{2}\sigma_n^z + h_{n,n+1}(\Delta).   
    \label{eq: nint II}
\end{align}
In both of the models, the bulk part is the same as the integrable XXZ spin chain, and the integrability is broken by the impurity term. In the model I, we take a similar form of the impurity Hamiltonian as the integrable impurity model at the leading order of $\xi$, except that we replace the anisotropic triple product $(\bm\sigma_{n-1},\bm\sigma_n,\bm\sigma_{n+1})_\Delta$ with the isotropic one $(\bm\sigma_{n-1},\bm\sigma_n,\bm\sigma_{n+1})_{\Delta=1}$. In the weak $\xi$ regime the difference between the integrable and non-integrable models arises only from the isotropic nature of the triple product, while in the strong $\xi$ regime higher order terms in $\xi$ in the integrable model may become dominant and give large deviation.
In the model II, we consider a perturbation by the local magnetic field in order to see how the current-induced magnetization found in the integrable model differs from the trivial magnetization induced by the magnetic field. We remark that both the triple-product term and the magnetic-field term have odd parity under the spin-flip transformation ($\bm\sigma_k \to -\bm\sigma_k$).
All the above impurity interactions we consider approach the non-impurity model as $\xi \to 0$. 


\subsection{Integrability of the $XXZ$ spin chain}
Now let us explain integrable aspects of the $XXZ$ spin chain and its impurity-doped deformation. The solvability of the $XXZ$ model is based on the existence of the commuting transfer matrices. The transfer matrix $T$ is constructed by taking the trace of the so-called monodromy matrix $\mathcal{T}_a: V_a \otimes \mathfrak{H} \to V_a \otimes \mathfrak{H}$, $V_a = \mathbb{C}^2$ over the auxiliary space $V_a$. The monodromy matrix consists of the Lax operators
\begin{align}
	&\mathcal{T}_a(\lambda) = L_{a,N}(\lambda) \cdots L_{a,1}(\lambda), \label{eq:monodromy} \\
	&
	L_{a,k}(\lambda) = 
	\begin{pmatrix}
		\sinh(\lambda \bm{1} + \eta S_k^z) & \sinh\eta \cdot S_k^- \\
		\sinh\eta \cdot S_k^+ & \sinh(\lambda \bm{1} - \eta S_k^z)
	\end{pmatrix}_a, \label{eq:Lax}
\end{align}
with the $U_q(\mathfrak{sl}_2)$ spin operators 
$S^+, S^-,e^{\pm \eta S^z}$
of spin-$1/2$. Each element in the Lax operator provides the Boltzmann weights of the six-vertex model parametrized by the complex parameter $\lambda \in \mathbb{C}$. 
The commuting property of the transfer matrix is based on the existence of invertible $\check{R}$-matrix which satisfies the fundamental commutation relation (FCR)
\begin{equation} \label{eq:RLL}
	\check{R}_{a,b}(\lambda - \mu) L_{a,k}(\lambda) L_{b,k}(\mu)
	= L_{a,k}(\mu) L_{b,k}(\lambda) \check{R}_{a,b}(\lambda - \mu). 
\end{equation}
For the above choice of the Lax operators, the invertible $\check{R}$-matrix which solves the $RLL$-relation indeed exists with the following form 
\begin{equation}
	\check{R}_{a,b}(\lambda) = 
	\begin{pmatrix}
		1 & 0 & 0 & 0 \\
		0 & \dfrac{\sinh\eta}{\sinh(\lambda + \eta)} & \dfrac{\sinh\lambda}{\sinh(\lambda + \eta)} & 0 \\
		0 & \dfrac{\sinh\lambda}{\sinh(\lambda + \eta)} & \dfrac{\sinh\eta}{\sinh(\lambda + \eta)} & 0 \\
		0 & 0 & 0 & 1
	\end{pmatrix}. 
\end{equation}
%
%
Using the FCR repeatedly, one obtains that the transfer matrices are commuting for any $\lambda$ due to invertibility of the $\check{R}$-matrix and the cyclic property of the trace 
\begin{align}
	&[T(\lambda),\,T(\mu)] = 0, \\
	&T(\lambda) = {\rm tr}_a \mathcal{T}_a(\lambda), \qquad  
	T(\mu) = {\rm tr}_b \mathcal{T}_b(\mu). 
\end{align}

The Hamiltonian of the $XXZ$ spin chain is obtained through the $\lambda$-expansion of the transfer matrix 
\begin{equation}
	\frac{1}{i} \log T(\lambda) = P + \frac{2(\lambda - \frac{\eta}{2})}{\sinh\eta} H_{XXZ} + \sum_{r=2}^{\infty} \frac{(\lambda - \frac{\eta}{2})^r}{r!} Q_r + {\rm const.}  
\end{equation}
%
%
Remind that the anisotropy of the model is parametrized by 
$\eta$
through the relation 
$\Delta = \cosh\eta$
. Note that the operators $Q_r$, which are operators with local but longer interactions than Hamiltonian, are all commuting due to the commuting property of the transfer matrices. 

The famous Yang-Baxter equation 
\begin{equation}
	R_{12} R_{13} R_{23} = R_{23} R_{13} R_{12}
\end{equation}
is a general interpretation of the fundamental commutation relation \eqref{eq:RLL}. In this interpretation, the Lax operator is obtained through the evaluation representation of the universal $\mathcal{R}$-matrix. For instance, the quasilocal charges~\cite{bib:LCN17, bib:INWCEP15, bib:IMPZ16, bib:M20, bib:SPA09} of the $XXZ$ model are constructed by choosing the auxiliary space $V_a$ in the spin-$s$ representation. When $s$ takes an integer or half-integer value, the conserved charges are spin-flip symmetric. On the other hand, if $s$ is a complex number, spin-flip asymmetric elements emerge in the conserved charges. 
The Lax operator looks the same as \eqref{eq:Lax} for any choice of the auxiliary space 
\begin{align}
	L_{k,a}(\lambda,s) = \begin{pmatrix}
		\sinh(\lambda \bm{1} + \eta S_a^z) & \sinh \eta \cdot S_a^- \\
		\sinh \eta \cdot S_a^+ & \sinh(\lambda \bm{1} - \eta S_a^z)
	\end{pmatrix}_k
\end{align}
%
%
but by replacing 
$S^+, S^-, e^{\pm \eta S^z}$
 with the $U_q(\mathfrak{sl}_2)$ spin operators of spin-$s$. 
They satisfy the commutation relations given by 
\begin{align}
    &[S^+,\,S^-] = \frac{q^{2S^z} - q^{-2S^z}}{q - q^{-1}}, \qquad
    q^{S^z} S^{\pm} q^{-S^z} = q^{\pm 1} S^{\pm}, 
\end{align}
in which the anisotropy of the model $\eta$ is encoded in the $q$ of $U_q(\mathfrak{sl}_2)$ via $q = e^{\eta}$. Note that the irreducible two-dimensional representations of $S^{\pm}$ coincide with those of the $\mathfrak{sl}_2$ generators. 
The explicit representations are given in \eqref{eq:Uq-integer} for integer or half-integer $s$, while the spin operators with complex spin $s$ have generically half-infinite dimensional representations with the highest weight $s$: 
\begin{align} \label{eq:Uq_inf}
	&e^{\pm \eta S_a^z} = \sum_{m=0}^{\infty} e^{\pm \eta (s-m)} |m \rangle \langle m|, \\
	&S_a^+ = \sum_{m=0}^{\infty} \frac{\sinh(\eta(m+1))}{\sinh\eta} |m \rangle \langle m+1|, \quad
	S_a^- = \sum_{m=0}^{\infty} \frac{\sinh(\eta(2s-m))}{\sinh\eta} |m+1 \rangle \langle m|. 
\end{align}
Only when the anisotropy is $\eta = i\gamma = i\pi k/l$ with coprime $k$ and $l$, we obtain finite-dimensional representations: 
\begin{align} \label{eq:Uq_finite}
	&e^{\pm i\gamma S_a^z} = \sum_{m=0}^{l-1} e^{\pm i\gamma (s-m)} |m \rangle \langle m|, \\
	&S_a^+ = \sum_{m=0}^{l-2} \frac{\sin(\gamma(m+1))}{\sin\gamma} |m \rangle \langle m+1|, \quad
	S_a^- = \sum_{m=0}^{l-2} \frac{\sin(\gamma(2s-m))}{\sin\gamma} |m+1 \rangle \langle m|. 
\end{align}
%
%

This is the story how to obtain the integrable Hamiltonian of the $XXZ$ type without impurity. In order to dope the spin-$1/2$ impurity at the $n$th site, use the monodromy matrix \eqref{eq:monodromy} whose $n$th Lax operator is replaced with $L(\lambda + \xi)$: 
\begin{equation} \label{eq:inhomo_monodromy}
	\mathcal{T}_a(\lambda;\xi) = L_{a,N}(\lambda) \cdots L_{a,n}(\lambda + \xi) \cdots L_{a,1}(\lambda), 
\end{equation}
where $\xi$ is the inhomogeneity parameter representing the effect of impurity. Similarly, the higher-spin impurity model is obtained from the monodromy matrix with the fused Lax operator at the $n$th site: 
\begin{equation} \label{eq:highers_monodromy}
	\mathcal{T}^{(\ell)}_a(\lambda) = L_{a,N}(\lambda) \cdots L^{(\ell)}_{a,n}(\lambda) \cdots L_{a,1}(\lambda), 
\end{equation}
where the Lax operator $L^{(\ell)}_{a,n}$ acts on $V_a \otimes \mathfrak{h}_n$, $\mathfrak{h}_n = \mathbb{C}^{2\ell+1}$. The deformations listed above do not break integrability of the model, since the monodromy matrices \eqref{eq:inhomo_monodromy} and \eqref{eq:highers_monodromy} also produce the commuting transfer matrices. The Hamiltonians constructed from these monodromy matrices contain the three-body interaction around the impurity, which is exactly what we are considering \eqref{eq:int-1a}, \eqref{eq:int-1b}, and \eqref{eq:int-2}.

\section{Steady state of the dissipative $XXZ$ model with impurity}
\subsection{Matrix product steady state}

In the series of works~\cite{bib:IZ14, bib:P11-2, bib:P11, bib:PIP13}, it has been found that the steady state of \eqref{eq:LB} can be exactly derived in the matrix product form in the absence of impurity. The key relation is {\it the Sutherland relation}: 
\begin{equation} \label{eq:Sutherland}
	[\check{R}'_{j,j+1}(0), \,L_{j,a}(\lambda,s) L_{j+1,a}(\lambda,s)] 
	= \sinh\eta \Big( L_{j,a}(\lambda,s) L'_{j+1,a}(\lambda,s) - L'_{j,a}(\lambda,s) L_{j+1,a}(\lambda,s) \Big)
\end{equation}
%
%
which are obtained as a consequence of the FCR. In the derivation of the Sutherland relation, the known fact that the local bulk Hamiltonian of the $XXZ$ model is obtained from the $\check{R}$-matrix at $\lambda = 0$ 
\begin{align}
	&\check{R}'_{j,j+1}(0) = 2 h_{j,j+1} + \frac{1}{2} \cosh\eta 
\end{align}
%
%
is used.  

The same Sutherland relation \eqref{eq:Sutherland} holds for the impurity-doped $XXZ$ model except for neighborhoods of the impurity. Around the impurity, we instead have the inhomogeneously deformed Sutherland relation: 
\begin{equation} \label{eq:Sutherland_Himp}
\begin{split} 
	[\check{R}^{-1}_{n-1,n}(\xi) \check{R}'_{n-1,n}(\xi) + \check{R}^{-1}_{n-1,n}(\xi) \check{R}'_{n,n+1}(0) \check{R}_{n-1,n}(\xi),\,
	L_{n-1,a}(\lambda,s) L_{n,a}(\lambda + \xi,s) L_{n+1,a}(\lambda,s)] \nonumber\\
	= \sinh\eta \Big( L_{n-1,a}(\lambda,s) L_{n,a}(\lambda + \xi,s) L'_{n+1,a}(\lambda,s) - L'_{n-1,a}(\lambda,s) L_{n,a}(\lambda + \xi,s) L_{n+1,a}(\lambda,s) \Big), 
\end{split}
\end{equation}
%
%
where the local interaction with impurity is proportional to the combination of $\check{R}$-operators appeared in the commutator: 
\begin{equation}
    2h_{n-1,n,n+1} = \check{R}^{-1}_{n-1,n}(\xi) \check{R}'_{n-1,n}(\xi) + \check{R}^{-1}_{n-1,n}(\xi) \check{R}'_{n,n+1}(0) \check{R}_{n-1,n}(\xi)
\end{equation}
Therefore, it is reasonable to assume that the steady state density matrix $\rho_{\infty}$ of the impurity-doped $XXZ$ model is also written in the matrix product form: 
\begin{align} \label{eq:SS}
	&\rho_{\infty} = \Omega_{N}^{\dag} \Omega_N / {\rm tr}(\Omega_{N}^{\dag} \Omega_N), \\
	&\Omega_N = {_a}\langle W_{\rm L} | L_{1,a}(\lambda,s) \cdots L_{n,a}(\lambda + \xi,s) \cdots L_{N,a}(\lambda,s) |W_{\rm R} \rangle_a, \\
	&\Omega_N^{\dag} = {_a}\langle W_{\rm L} | \overline{L_{1,a}^{\rm T}(\lambda,s)} \cdots \overline{L_{n,a}^{\rm T}(\lambda + \xi,s)} \cdots \overline{L_{N,a}^{\rm T}(\lambda,s)} |W_{\rm R} \rangle_a. 
\end{align}
Here the superscript ${\rm T}$ represents partial transpose with respect to the physical space. $|W_{\rm R} \rangle_a$ is a certain vector in the auxiliary space $V_a$ and ${_a}\langle W_{\rm L}|$ is in its adjoint space $V_a^*$. Note that the decomposition of the density matrix $\rho_{\infty}$ into the amplitude operators $\Omega_N$ is always possible due to positive definiteness of the density matrix. Remind that the FCR holds for any choice of the auxiliary space of the Lax operators. Therefore, we let the auxiliary space in an arbitrary spin-$s$ representation, which must be fixed in order to be compatible with the steady state condition at the boundaries. 

Repeatedly using the Sutherland relation \eqref{eq:Sutherland} and its inhomogeneous deformation \eqref{eq:Sutherland_Himp}, the non-commuting parts in the commutator of \eqref{eq:LB} remain only at the boundaries, which can be canceled by the dissipation terms. By choosing the boundary vectors as the highest-weight vectors ${_a}\langle W_{\rm L}| = {_a}\langle 0|,\,|W_{\rm R} \rangle_a = |0 \rangle_a$, the matrix product state \eqref{eq:SS} satisfies the steady state condition $\mathcal{L}(\rho) = 0$ at $\lambda = i\pi/2$, provided that the spin parameter $s$ is connected to the dissipation rates through the relation 
$\varepsilon_{\rm L} = \varepsilon_{\rm R} = -2i \sinh\eta \tanh(\eta s)$. 
Since $s$ is allowed to take any complex number, the dissipation rates can be any, provided that $\varepsilon_{\rm L} = \varepsilon_{\rm R}$. Note that, in order for the dissipation rate to be real, the spin parameter $s$ must be pure imaginary. 

The steady state of the $XXZ$ model with the spin-$\ell/2$ impurity is similarly obtained through the Sutherland relation \eqref{eq:Sutherland} and its higher-spin deformation:
\begin{equation}
\begin{split}
    &[ (R_{n-1,n}^{(\ell,\frac{1}{2})}(0))^{-1} (\check{R}_{n,n+1}(0))' R_{n-1,n}^{(\frac{1}{2},\ell)}(0) + (R_{n-1,n}^{(\ell,\frac{1}{2})}(0))^{-1} \check{R}_{n,n+1}(0) (R_{n-1,n}^{(\frac{1}{2},\ell)}(0))',\,L_{n-1,a}(\lambda,s) L_{n,a}^{(\ell)}(\lambda,s) L_{n+1,0}(\lambda,s)] \\
    &= i \sin\gamma \Big( L_{n-1,a}(\lambda,s) L_{n,a}^{(\ell)}(\lambda,s) L'_{n+1,a}(\lambda,s) - L'_{n-1,a}(\lambda,s) L_{n,a}^{(\ell)}(\lambda,s) L_{n+1,a}(\lambda,s) \Big). 
\end{split}
\end{equation}
$L_{n,a}^{(\ell)}$ is the fused Lax operator with $(2\ell + 1)$-dimensional physical space and $R_{a,b}^{(\frac{1}{2},\ell)} \in {\rm End}(\mathbb{C}^{2} \otimes \mathbb{C}^{2\ell+1})$ is the fused $R$-matrix~\cite{bib:KRS81}. The local Hamiltonian around the impurity is proportional to the product of the $\check{R}$-matrices in the commutator: 
\begin{equation}
    2h_{n-1,n,n+1} = (R_{n-1,n}^{(1,\frac{1}{2})}(0))^{-1} (\check{R}_{n,n+1}(0))' R_{n-1,n}^{(\frac{1}{2},1)}(0) + (R_{n-1,n}^{(1,\frac{1}{2})}(0))^{-1} \check{R}_{n,n+1}(0) (R_{n-1,n}^{(\frac{1}{2},1)}(0))'. 
\end{equation}
The amplitude operator for the spin-$\ell/2$-impurity-doped model is then given by the matrix product form: 
\begin{align}
	&\Omega_N = {_a}\langle 0 | L_{1,a}(\tfrac{i\pi}{2},s) \cdots L^{(\ell)}_{n,a}(\tfrac{i\pi}{2},s) \cdots L_{N,a}(\tfrac{i\pi}{2},s) |0 \rangle_a. 
\end{align}
under the condition that the spin parameter $s$ is determined by the dissipation rate via 
$\varepsilon_{\rm L} = \varepsilon_{\rm R} = -2i \sinh\eta \tanh(\eta s)$. 

\subsection{Physical quantities in the steady state}
Once the steady state is given in the matrix product form, expectation values of physical quantities are also expressed in the matrix product forms. In the rest of the paper, we focus our discussion on the spin-$1/2$ impurity [Eqs.~(\ref{eq:int-1a}) and (\ref{eq:int-1b})]. We are especially interested in the magnetization profile
\begin{equation}
	 \langle \sigma_k^z \rangle = {\rm tr}(\rho_{\infty} \sigma_k^z),
\end{equation}
and the spin current,
\begin{equation}
    \langle j_{k,k+1} \rangle = {\rm tr}(\rho_{\infty} j_{k,k+1}),
\end{equation}
which explicitly depend on the site $k$ due to the presence of the impurity. For the gapless regime $|\Delta| \leq 1$, we have 
\begin{align}
	&j_{k,k+1} = 4i (\sigma_k^+ \sigma_{k+1}^- - \sigma_k^- \sigma_{k+1}^+) \qquad (k \neq n-1,n), \\
	&j_{n-1,n} = 4i \frac{1 - \Delta^2}{1 - \Delta^2 + \sinh^2\xi} \cosh\xi (\sigma_{n-1}^+ \sigma_{n}^- - \sigma_{n-1}^- \sigma_{n}^+) \nonumber \\
		&\hspace{10mm}+ 4 \frac{1}{1 - \Delta^2 + \sinh^2\xi} \Big( i \Delta \sinh^2\xi (\sigma_{n-1}^+ \sigma_{n+1}^- - \sigma_{n-1}^- \sigma_{n+1}^+) \nonumber \\
		&\hspace{44mm}- \Delta \sqrt{1 - \Delta^2} \sinh\xi (\sigma_{n-1}^+ \sigma_n^- + \sigma_{n-1}^- \sigma_n^+) \sigma_{n+1}^z \nonumber \\
		&\hspace{44mm}+ \sqrt{1 - \Delta^2} \sinh\xi \cosh\xi (\sigma_{n-1}^+ \sigma_n^z \sigma_{n+1}^- + \sigma_{n-1}^- \sigma_n^z \sigma_{n+1}^+) \Big), \label{eq:current_L}\\
	&j_{n,n+1} = 4i \frac{1 - \Delta^2}{1 - \Delta^2 + \sinh^2\xi} \cosh\xi (\sigma_{n}^+ \sigma_{n+1}^- - \sigma_{n}^- \sigma_{n+1}^+) \nonumber \\
		&\hspace{10mm}+ 4 \frac{1}{1 - \Delta^2 + \sinh^2\xi} \Big( i \Delta \sinh^2\xi (\sigma_{n-1}^+ \sigma_{n+1}^- - \sigma_{n-1}^- \sigma_{n+1}^+) \nonumber \\
		&\hspace{44mm}- \Delta \sqrt{1 - \Delta^2} \sinh\xi \cdot \sigma_{n-1}^z (\sigma_{n}^+ \sigma_{n+1}^- + \sigma_{n}^- \sigma_{n+1}^+) \nonumber \\
		&\hspace{44mm}+ \sqrt{1 - \Delta^2} \sinh\xi \cosh\xi (\sigma_{n-1}^+ \sigma_n^z \sigma_{n+1}^- + \sigma_{n-1}^- \sigma_n^z \sigma_{n+1}^+) \Big), \label{eq:current_R}
\end{align}
where $\sigma^{\pm} = (\sigma^x \pm i \sigma^y) / 2$. The spin current in the gapped regime $
|\Delta| > 1$ is similarly obtained. Note that the definition of the spin current around the impurity must be modified from the one away from the impurity as \eqref{eq:current_L} and \eqref{eq:current_R} in order to satisfy the continuity equation (see Appendix \ref{sec:spin_currents}). 

For practical calculations, let us introduce {\it the double Lax operator} $\mathbb{L}_{k,a,b}$~\cite{bib:P15}, which acts in the tensor product of a quantum space and two auxiliary spaces $\mathfrak{h}_k \otimes V_a \otimes V_b$:
\begin{equation}
	\mathbb{L}_{k,a,b}(\lambda,s) = L^{\dag}_{k,a}(\lambda,s) L_{k,b}(\lambda,s). 
\end{equation}
From now on, we fix the spin parameter $s$ of the double Lax operator $\mathbb{L}$ at the steady state point $2i \sin\gamma \tan(\gamma s) = \varepsilon$ unless explicitly noted.  
The steady state density matrix $\rho_{\infty} = \Omega_N^{\dag} \Omega_N / {\rm tr}(\Omega_N^{\dag} \Omega_N)$ is expressed as the product of the double Lax operators via 
\begin{align} \label{eq:MPS}
	\Omega_N^{\dag} \Omega_N &= {_a}\langle 0| \otimes {_b}\langle 0| \prod_{x=1}^N \mathbb{L}_{x,a,b}(\lambda^s_x)  |0 \rangle_a \otimes |0 \rangle_b, 
\end{align}
where $\{\lambda^s_x\}$ is a set of spectral parameters given by $(\lambda^s_1,\dots,\lambda^s_n,\dots,\lambda^s_N) = (i\pi/2,\dots,i\pi/2+\xi,\dots,i\pi/2)$. 

It is also useful to rewrite the double Lax operator element by the elements with respect to Pauli matrices
\begin{equation}
	\mathbb{L}_{k,a,b}(\lambda) = \sum_{\alpha \in \{0,z,+,-\}} \sigma_k^{\alpha} \otimes \mathbb{L}_{a,b}^{\alpha}(\lambda) 
	= \begin{pmatrix} \mathbb{L}_{a,b}^0(\lambda) + \mathbb{L}_{a,b}^z(\lambda) & \mathbb{L}_{a,b}^+(\lambda) \\ \mathbb{L}_{a,b}^-(\lambda) & \mathbb{L}_{a,b}^0(\lambda) - \mathbb{L}_{a,b}^z(\lambda) \end{pmatrix}_k. 
\end{equation}
Then, the normalization constant of \eqref{eq:MPS}, which we call {\it the partition function}, is expressed simply by the $0$-element of the double Lax operator: 
\begin{equation}
\begin{split}
	 Z_N := {\rm tr}(\Omega_N^{\dag} \Omega_N)
	= {_a}\langle 0| \otimes {_b}\langle 0| \prod_{x=1}^N \mathbb{L}_{a,b}^0(\lambda^s_x) |0 \rangle_a \otimes |0 \rangle_b, 
\end{split}
\end{equation}
since only the tensor product of $\sigma^0$ survives under the trace over the physical spaces. 

Recall that the expectation value of a physical quantity $X$ on the steady state is evaluated by $\langle X \rangle = {\rm tr}(\rho_{\infty} X)$. Noting that the trace over the physical space kills any tensor product operator which includes the Pauli matrices, we obtain the matrix product forms of the local magnetization and the local spin current in the steady state as
\begin{align}
	&\langle \sigma_k^z \rangle 
	= \frac{{_a}\langle 0| \otimes {_b}\langle 0| \prod_{x=1}^{k-1} \mathbb{L}_{a,b}^0(\lambda^s_x) \mathbb{L}_{a,b}^z(\lambda^s_k) \prod_{x=k+1}^N \mathbb{L}_{a,b}^0(\lambda^s_x) |0 \rangle_a \otimes |0 \rangle_b}
 {{_a}\langle 0| \otimes {_b}\langle 0| \prod_{x=1}^{N} \mathbb{L}_{a,b}^0(\lambda^s_x) |0 \rangle_a \otimes |0 \rangle_b}, 
 \label{eq: S_k^z}
	\\
	&\langle j_{k,k+1} \rangle = 4i \frac{{_a}\langle 0| \otimes {_b}\langle 0| \prod_{x=1}^{k-1} \mathbb{L}_{a,b}^0(\lambda^s_x) [\mathbb{L}_{a,b}^-(\lambda^s_k),\,\mathbb{L}_{a,b}^+(\lambda^s_{k+1})] \prod_{x=k+2}^N \mathbb{L}_{a,b}^0(\lambda^s_x) |0 \rangle_a \otimes |0 \rangle_b}
 {{_a}\langle 0| \otimes {_b}\langle 0| \prod_{x=1}^{N} \mathbb{L}_{a,b}^0(\lambda^s_x) |0 \rangle_a \otimes |0 \rangle_b}. \label{eq:current}
\end{align}
The matrix product expressions for expectation values of general $\Lambda(\in \mathbb{N})$-local operators are found in Appendix \ref{sec:MPforms}. 

We obtain the expectation values of physical quantities that depend on the position of the impurity $n$. Indeed, magnetization loss/gain emerges at the impurity site (Fig.~\ref{fig:S^z at t=40}). On the other hand, the spin current is constant throughout the chain, which follows from the continuity equation. 

In order to simplify the discussion, let us focus on the gapless regime $\Delta=\cos\gamma\le 1$ with $\gamma = \pi k/l$, in which $k$ and $l(>k)$ are coprime. In this case, the $U_q(\mathfrak{sl}_2)$ spin operators $S^+, S^-, e^{\pm i\gamma S^z}$ have finite $l$-dimensional representations, and subsequently, the double Lax operator element $\mathbb{L}_{a,b}^0$ has an invariant subspace spanned by finite number of vectors $\mathcal{V}_{\rm sub} := \{|m \rangle_a \otimes |m \rangle_b\}_{m = 0,\dots,l-1}$~\cite{bib:P15}. The expression for $\mathbb{L}_{a,b}^0$ restricted to this subspace is 
\begin{align} \label{eq:L0}
    \bm{L}_{a,b}^{0}(\lambda) := \mathbb{L}_{a,b}^0(\lambda) |_{\mathcal{V}_{\rm sub}}
    &= \sum_{m=0}^{l-1}|\sinh\lambda \cdot \cos(\gamma(s-m))|^2\, |m,m \rangle \langle m,m| \nonumber\\
	&+ \sum_{m=0}^{l-1} |\cosh\lambda \cdot i\sin(\gamma(s-m))|^2\, |m,m \rangle \langle m,m| \nonumber\\
	&+ \frac{1}{2} \sum_{m=0}^{l-2} |i\sin(\gamma (2s-m))|^2\, |m+1,m+1 \rangle \langle m,m| \nonumber\\
	&+ \frac{1}{2} \sum_{m=0}^{l-2} |i\sin(\gamma(m+1))|^2\, |m,m \rangle \langle m+1,m+1|. 
\end{align}
Here we simply denote $|m \rangle_a \otimes |m \rangle_b$ by $|m,m \rangle$ etc. The commutator $[\mathbb{L}_{a,b}^+ \mathbb{L}_{a,b}^-,\, \mathbb{L}_{a,b}^- \mathbb{L}_{a,b}^+]$ also have the same invariant subspace, whose expressions in this subspace are 
\begin{align} \label{eq:L+L-}
	[\bm{L}_{a,b}^+(\lambda),\, \bm{L}_{a,b}^-(\lambda)]  
     &:= [\mathbb{L}_{a,b}^+(\lambda),\, \mathbb{L}_{a,b}^-(\lambda)]\Big|_{\mathcal{V}_{\rm sub}}
     \\
     &= i\sin\gamma \sum_{m=0}^{l-2} \sinh(\overline{\lambda} + \lambda -i\gamma (\overline{s} - s)) |i\sin(\gamma (2s - m))|^2 |m+1,m+1 \rangle \langle m,m| \nonumber \\
     &- i\sin\gamma \sum_{m=0}^{l-2} \sinh(\overline{\lambda} + \lambda + i\gamma (\overline{s} - s)) |i\sin(\gamma (m+1))|^2 |m,m \rangle \langle m+1,m+1| \nonumber \\
     &- \frac{i}{2}\sin\gamma \sum_{m=0}^{l-1} (\cosh(2\overline{\lambda}) - \cosh(2i\gamma (\overline{s}-m))) \sinh(2i\gamma (s-m)) |m,m \rangle \langle m,m| \nonumber \\
     &+ \frac{i}{2}\sin\gamma \sum_{m=0}^{l-1} \sinh(2i\gamma (\overline{s}-m)) (\cosh(2\lambda) - \cosh(2i\gamma (s-m))) |m,m \rangle \langle m,m|. 
\end{align}
By carefully comparing \eqref{eq:L0} with \eqref{eq:L+L-}, the relation 
\begin{equation}
	[\bm{L}_{a,b}^+(\tfrac{i\pi}{2}),\, \bm{L}_{a,b}^-(\tfrac{i\pi}{2})] = -i\zeta \bm{L}_{a,b}^0(\tfrac{i\pi}{2}), \qquad
	\zeta = \frac{\varepsilon \sin^2\gamma}{|\frac{1}{4} \varepsilon^2 - \sin^2\gamma|}  
\end{equation}
holds at $\lambda = i\pi/2$~\cite{bib:P15}. 
Using this relation, the spin current \eqref{eq:current} except for neighborhoods of the impurity ($k \neq n-1,n$) is straightforwardly expressed by the ratio of the partition functions: 
\begin{equation}
	\langle j_{k,k+1} \rangle = 4\zeta \frac{Z_{N-1}}{Z_N},  
\end{equation}
which is the quantum analog of the matrix product form of currents obtained for the asymmetric simple exclusion process~\cite{bib:DEHP93}. Moreover, the spin current is site-independent in the steady state due to the continuity equation, leading to 
\begin{equation}
    \langle j_{1,2} \rangle = \cdots = \langle j_{n-1,n} \rangle = \langle j_{n,n+1} \rangle = \cdots = \langle j_{N-1,N} \rangle = 4\zeta \frac{Z_{N-1}}{Z_N}. 
\end{equation}

\subsection{Thermodynamic properties}
In the thermodynamic limit $N \to \infty$, the number of $\mathbb{L}_{a,b}^{0}$ in the matrix product form of {\it local} physical quantities linearly grows with the system size $N$. The restricted Lax operator $\bm{L}_{a,b}^{0}(\frac{i\pi}{2})$ has the positive largest eigenvalue $\tau_0$ since it is finite dimensional with non-negative real elements. Besides, eigenvalues of $\bm{L}_{a,b}^{0}(\frac{i\pi}{2})$ are strictly contracting $1 > |\tau_0| > |\tau_1| \geq \dots \geq |\tau_{\ell-1}|$, if properly ordered, since $\bm{1} - \bm{L}_{a,b}^{0}(\frac{i\pi}{2})$ is positive definite. Therefore, the eigenvector of the largest eigenvalue $\tau_0$ gives the leading term. 
Thus, we have the large-$N$ behavior of the spin current as 
\begin{equation}
	\langle j_{k,k+1} \rangle = 4\zeta \tau_0^{-1}, 
\end{equation}
which shows that the presence of impurity does not affect the spin current in the thermodynamic limit. This seems to be the remarkable property of the system with bulk integrability in comparison with the numerical results of non-integrable cases, which shows that the spin current decreases as the coupling with impurity increases. Note that this expression is true for any $k$ due to the continuity equation of the spin current. 

On the other hand, the large-$N$ behavior of the magnetization is obtained by using the existence of such a basis that makes $\bm{L}_{a,b}^{0}$ symmetric and $\bm{L}_{a,b}^z$ anti-symmetric (Appendix \ref{sec:symmetrization}). Let $|\phi_m \rangle$ ({\it resp.} $\langle \psi_j|$) be a right ({\it resp.} left) eigenvector of $\bm{L}_{a,b}^0(\frac{i\pi}{2})$ for an eigenvalue $\tau_m$ ($m = 0,\dots,\ell-1$). We also write the eigenvalue of $\bm{L}_{a,b}^0(\frac{i\pi}{2} + \xi)$ to the eigenvector $|\phi_m \rangle$ by $\tau_m'$, since $\bm{L}_{a,b}^0(\frac{i\pi}{2} + \xi)$ is simultaneously diagonalizable with $\bm{L}_{a,b}^0(\frac{i\pi}{2})$.
Then we obtain that the magnetization asymptotically behaves as 
\begin{align}
	\langle \sigma^z_{k > n} \rangle = 
	\sum_{j=1}^{l-1} 
	\Bigg\{  
	&\frac{\tau'_j}{\tau'_0} \left( \frac{\tau_j}{\tau_0} \right)^{k-2} \tau_0^{-1} \frac{\langle 0,0| \phi_j \rangle}{\langle 0,0| \phi_0 \rangle} \langle \psi_j | \bm{L}_{a,b}^z(\tfrac{i\pi}{2}) | \phi_0 \rangle \frac{\langle \psi_0 | 0,0 \rangle}{\langle \psi_0 | 0,0 \rangle} \nonumber\\ 
	&+ \left( \frac{\tau_j}{\tau_0} \right)^{N-k} \tau_0^{-1} \frac{\langle 0,0| \phi_0 \rangle}{\langle 0,0| \phi_0 \rangle} \langle \psi_0 | \bm{L}_{a,b}^z(\tfrac{i\pi}{2}) | \phi_j \rangle \frac{\langle \psi_j |0,0 \rangle}{\langle \psi_0 |0,0 \rangle}
	\Bigg\} 
	\\
	\langle \sigma^z_{k < n} \rangle = 
	\sum_{j=1}^{l-1} 
	\Bigg\{  
	& \left( \frac{\tau_j}{\tau_0} \right)^{k-1} \tau_0^{-1} \frac{\langle 0,0| \phi_j \rangle}{\langle 0,0| \phi_0 \rangle} \langle \psi_j | \bm{L}_{a,b}^z(\tfrac{i\pi}{2}) | \phi_0 \rangle \frac{\langle \psi_0 | 0,0 \rangle}{\langle \psi_0 | 0,0 \rangle} \nonumber\\ 
	&+ \frac{\tau'_j}{\tau'_0} \left( \frac{\tau_j}{\tau_0} \right)^{N-k-1} \tau_0^{-1} \frac{\langle 0,0| \phi_0 \rangle}{\langle 0,0| \phi_0 \rangle} \langle \psi_0 | \bm{L}_{a,b}^z(\tfrac{i\pi}{2}) | \phi_j \rangle \frac{\langle \psi_j |0,0 \rangle}{\langle \psi_0 |0,0 \rangle}
	\Bigg\} 
\end{align}
at the non-impurity site, while 
\begin{align} \label{eq:asym_Sz}
	\langle \sigma^z_{k = n} \rangle =  
	\tau_0^{'-1} \frac{\langle 0,0| \phi_0 \rangle}{\langle 0,0| \phi_0 \rangle} &\langle \psi_0| \bm{L}_{a,b}^z(\tfrac{i\pi}{2} + \xi) | \phi_0 \rangle \frac{\langle \psi_0 |0,0 \rangle}{\langle \psi_0 |0,0 \rangle} \nonumber \\
	+ \sum_{j=1}^{l-1} 
	\Bigg\{  
	&\left( \frac{\tau_j}{\tau_0} \right)^{n-1} \tau_0^{'-1} \frac{\langle 0,0| \phi_j \rangle}{\langle 0,0| \phi_0 \rangle} \langle \psi_j | \bm{L}_{a,b}^z(\tfrac{i\pi}{2} + \xi) | \phi_0 \rangle \frac{\langle \psi_0 |0,0 \rangle}{\langle \psi_0 |0,0 \rangle} \nonumber\\ 
	&+ \left( \frac{\tau_j}{\tau_0} \right)^{N-n} \tau_0^{'-1} \frac{\langle 0,0| \phi_0 \rangle}{\langle 0,0| \phi_0 \rangle} \langle \psi_0 | \bm{L}_{a,b}^z(\tfrac{i\pi}{2} + \xi) | \phi_j \rangle \frac{\langle \psi_j |0,0 \rangle}{\langle \psi_0 |0,0 \rangle}
	\Bigg\}  
\end{align}
at the site of impurity. 
Due to the presence of the diagonal term (the first term) in $\langle \sigma^z_{k = n} \rangle$, the magnetization shows discontinuity at the site of impurity, proving the emergence of local magnetization in the thermodynamic limit.

\section{Numerical results}

In this section, we demonstrate the numerical results for the Lindblad master equation (\ref{eq:LB}) for the finite-size $XXZ$ spin chain with the spin-1/2 impurity (Eqs.~(\ref{eq:int-1a}) and (\ref{eq:int-1b})) and boundary dissipation. Our approach is based on the quantum-trajectory method \cite{bib:Carmichael93, bib:Daley14, big:DCM92} combined with the exact diagonalization.
We set the initial state to be the N\'{e}el state $|\uparrow\downarrow\uparrow\downarrow\cdots\rangle$. 
The ground state of the bulk Hamiltonian with the impurity $H_{XXZ}^{\rm imp}$ (\ref{eq: H imp}) under the open boundary condition has vanishing expectation values for the local magnetization, $\langle \sigma_k^z\rangle=0$, for arbitrary $\xi$.

\begin{figure}[t]
    \centering
    \includegraphics[width=7cm]{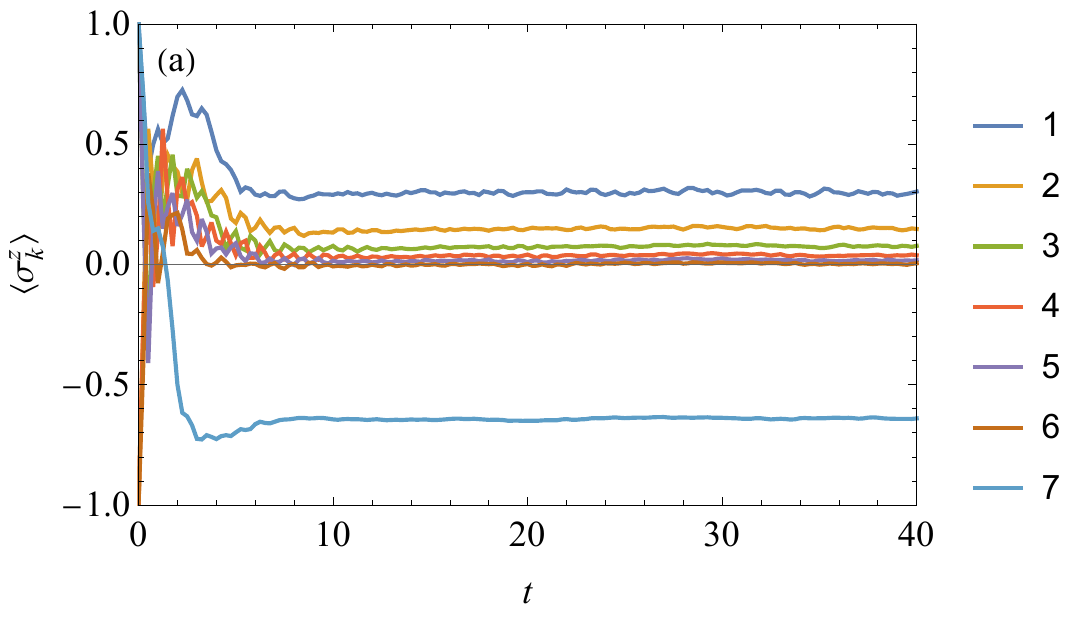}
    \includegraphics[width=7cm]{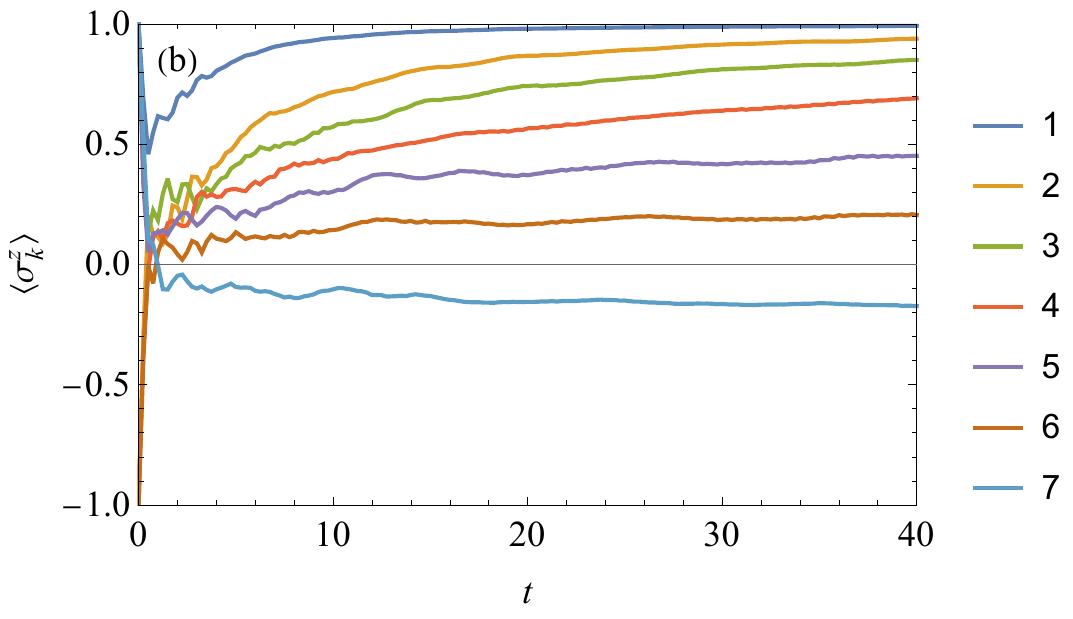}
    \caption{Time evolution of the local magnetization $\langle\sigma_k^z\rangle$ ($k=1,2,\dots,7$) obtained from the quantum trajectory method for the dissipative $XXZ$ model with (a) $\Delta=0.5$ and (b) $\Delta=1.5$ with the impurity inserted at the site $k=7$
    ($N=12, \varepsilon_{\rm R}=\varepsilon_{\rm L}=1, \xi=1$). The initial state is the N\'{e}el state. We take the average over 4000 trajectories.}
    \label{fig:S^z time evolution}
\end{figure}

In Fig.~\ref{fig:S^z time evolution}, we show the time evolution of the local magnetization $\langle\sigma_k^z\rangle$ ($k=1,2,\dots,7$) for the integrable dissipative $XXZ$ spin chain with $\Delta=0.5$ (Fig.~\ref{fig:S^z time evolution}(a)) and $\Delta=1.5$ (Fig.~\ref{fig:S^z time evolution}(b)) with the system size $N=12$. The impurity is inserted at the site $k=7$. For $\Delta=0.5$,
one can see that the local magnetization at the impurity site ($\langle\sigma_{k=7}^z\rangle$) rapidly grows to the negative value after the system arrives at the steady state around $t\sim 10$, while the other sites for $k\le 6$ have relatively small positive magnetization (except near the boundary). Due to the injection/ejection of spins at the boundaries, the spin current flows steadily through the system, which does not suppress the local magnetization but rather enhances the magnetization at the impurity site. This {\it current-induced magnetization} is a unique phenomenon in out-of-equilibrium spin chains with impurities, and is in stark contrast to our naive expectation that magnetization would be suppressed due to heating induced by current flow. Let us remark that the local magnetization vanishes at the ground state of the bulk Hamiltonian without dissipators. For $\Delta=1.5$, the system shows the development of local magnetization throughout the bulk, and the impurity does not particularly enhance the magnetization near the impurity site. 
The difference of the impurity effect between the case of $\Delta=0.5$ and $\Delta=1.5$ may come from the difference of the magnitude of the current. In the gapless regime ($\Delta <1$), the magnitude of the steady-state current is in the order of 1. Hence the transported spin can be trapped around the impurity site, inducing local magnetization. On the other hand, in the gapped regime ($\Delta >1$) the steady-state current is exponentially small as the system size grows \cite{bib:P11-2}. Hence there is no transported spins that can be trapped at the impurity site, and there is no current-induced magnetization.

\begin{figure}[t]
    \centering
    \includegraphics[width=7cm]{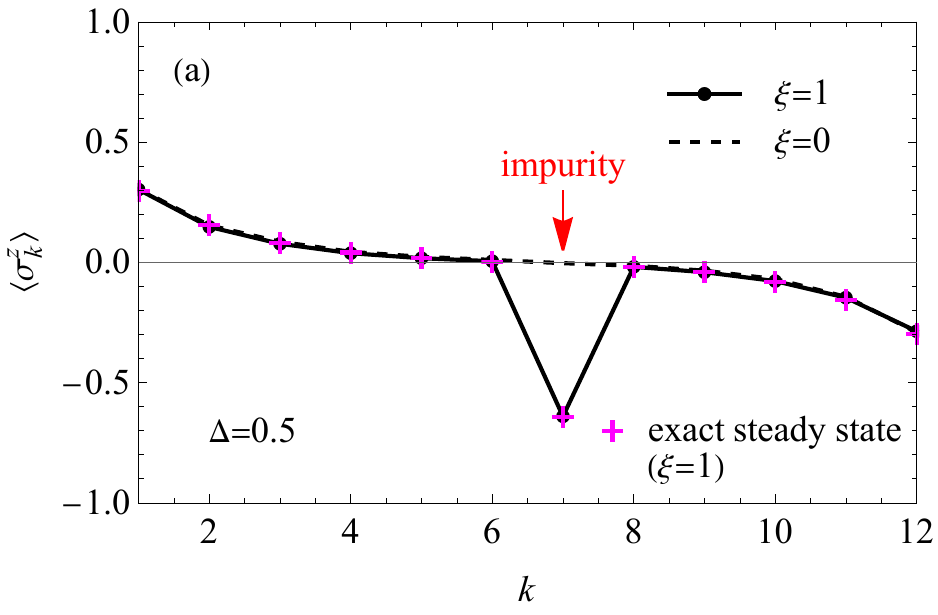}
    \includegraphics[width=7cm]{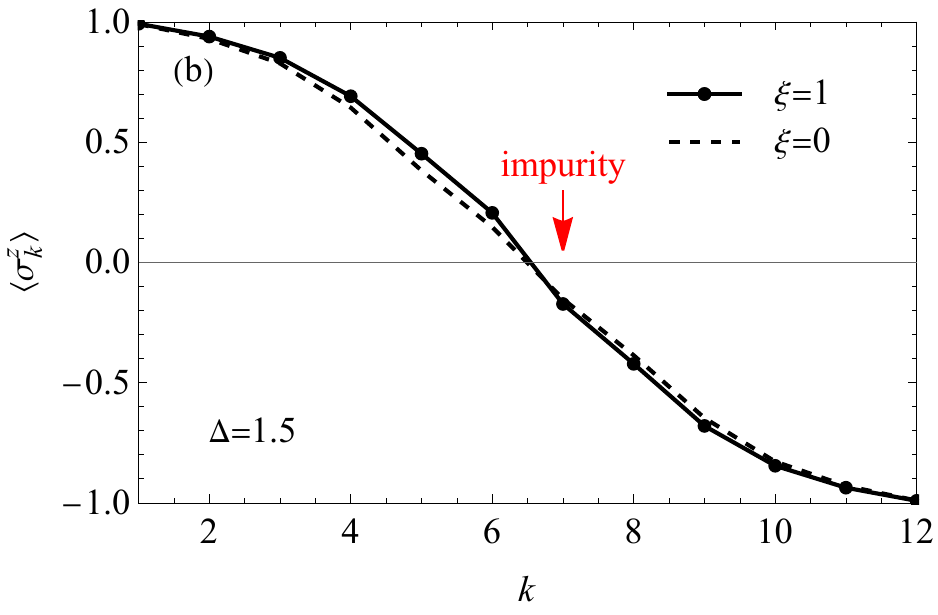}
    \caption{(Solid line) Site dependence of the local magnetization $\langle\sigma_k^z\rangle$ at $t=40$ for the dissipative $XXZ$ model
    with (a) $\Delta=0.5$ and (b) $\Delta=1.5$ with the impurity inserted at the site $k=7$
    ($N=12, \varepsilon_{\rm R}=\varepsilon_{\rm L}=1, \xi=1$) obtained by the quantum trajectory method. (Dashed line) The corresponding $\langle\sigma_k^z\rangle$ for the dissipative $XXZ$ model without an impurity ($\xi=0$). The initial state is the N\'{e}el state. We take the average over 4000 trajectories. (Cross marks) Exact results for the steady state solution (Eq.~(\ref{eq: S_k^z})) of the dissipative $XXZ$ model with the impurity ($\xi=1$).}
    \label{fig:S^z at t=40}
\end{figure}

In Fig.~\ref{fig:S^z at t=40}, we plot the real-space profile of the local magnetization $\langle\sigma_k^z\rangle$ for the integrable $XXZ$ spin chain with $\Delta=0.5$ (Fig.~\ref{fig:S^z at t=40}(a)) and $\Delta=1.5$ (Fig.~\ref{fig:S^z at t=40}(b)) with and without the impurity ($\xi=1$ and $\xi=0$) at $t=40$, where the system nearly reaches the steady state. In the case of $\Delta=0.5$ (Fig.~\ref{fig:S^z at t=40}(a)), the result for $\xi=0$ shows that the local magnetization appears only near the boundaries and decays quickly inside the bulk. This is consistent with the previous study \cite{bib:P11-2}. As we include the impurity ($\xi=1$), the magnetization at the impurity site ($k=7$) grows in the steady state, while the rest of the sites has almost the same magnetization as in the case of $\xi=0$. Thus, the current-induced magnetization is localized around the impurity site, without affecting the bulk spin configuration. In the case of $\Delta=1.5$ (Fig.~\ref{fig:S^z at t=40}(b)), where the bulk system is in the gapped regime, the effect of the impurity is much suppressed as compared to the case of $\Delta=0.5$. In fact, the configuration of the magnetization remains almost the same as the one without the impurity ($\xi=0$).
The results agree well with the exact solution (Eq.~(\ref{eq: S_k^z})) for the steady state of the finite-size chain (cross marks in Fig.~\ref{fig:S^z at t=40}(a)). Note that the exact solution is numerically not available for $\Delta>1$ since the dimension of the matrices in the matrix product form of the steady state becomes infinite 
\eqref{eq:Uq_inf} 
in those cases.

Next, we compare the results between the integrable and non-integrable impurity models. For the latter, we take the Hamiltonian (\ref{eq: H imp}) with $h_{n-1,n,n+1}=h_{n-1,n,n+1}^{\rm nint(I)}$ (\ref{eq: nint I}) and $h_{n-1,n,n+1}=h_{n-1,n,n+1}^{\rm nint(II)}$ (\ref{eq: nint II}). The entire form of the Hamiltonian is
\begin{align}
    H^{\rm nint(I)}
    &=
    \sum_{j=1}^{N-1} h_{j,j+1}(\Delta)
    +\frac{\xi}{2}\frac{1}{\sqrt{|1-\Delta^2|}}
    (\bm\sigma_{n-1}, \bm\sigma_n, \bm\sigma_{n+1})_{\Delta=1}
\end{align}
for the model I, and
\begin{align}
    H^{\rm nint(II)}
    &=
    \sum_{j=1}^{N-1} h_{j,j+1}(\Delta)
    +\frac{\xi}{2}\sigma_n^z
\end{align}
for the model II.

\begin{figure}[htbp]
    \centering
    \includegraphics[width=7.2cm]{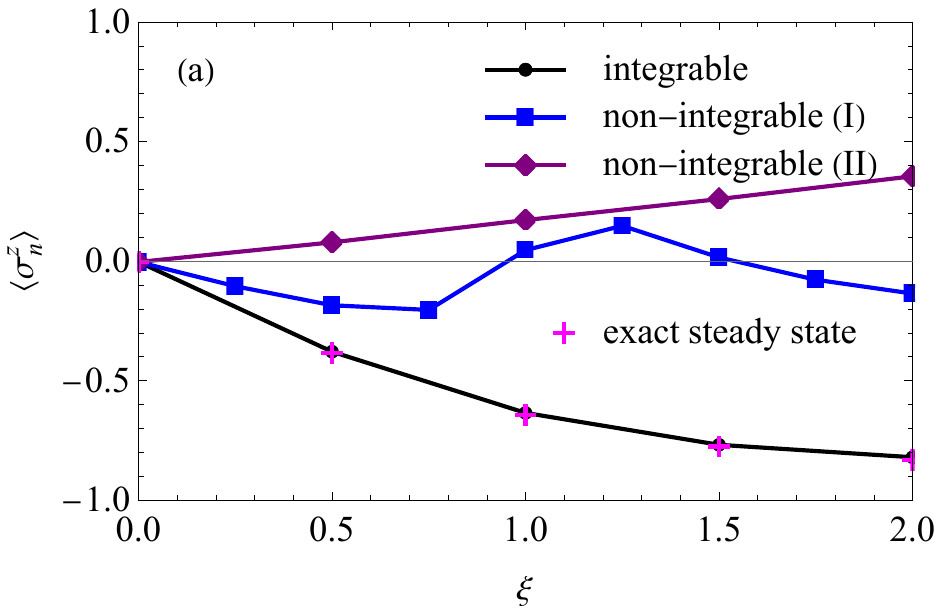}
    \includegraphics[width=7cm]{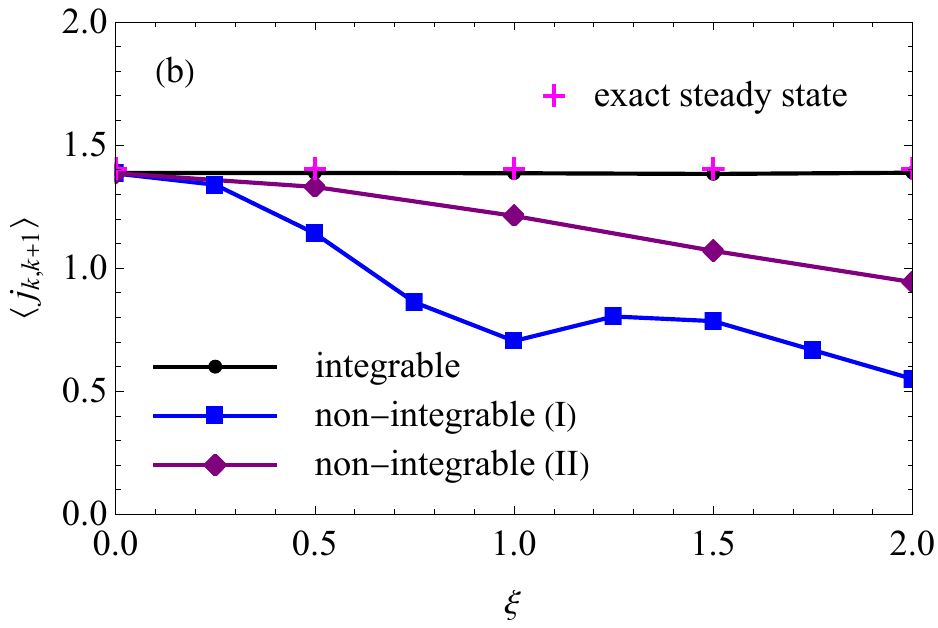}
    \caption{The local magnetization at the impurity site $\langle\sigma_n^z\rangle$ and (b) the spin current $\langle j_{k,k+1}\rangle$ at $t=80$ as a function of the impurity strength $\xi$ for the integrable model and non-integrable models I and II ($N=12, n=7, \Delta=0.5, \varepsilon_{\rm R}=\varepsilon_{\rm L}=1$) obtained by the quantum trajectory method. The initial state is the N\'{e}el state. We take the average over 4000 trajectories. (Cross marks) Exact results for the steady state solution (Eqs.~(\ref{eq: S_k^z}) and (\ref{eq:current})) of the integrable model.}
    \label{fig:S^z xi}
\end{figure}
\begin{figure}[htbp]
    \centering
    \includegraphics[width=7.2cm]{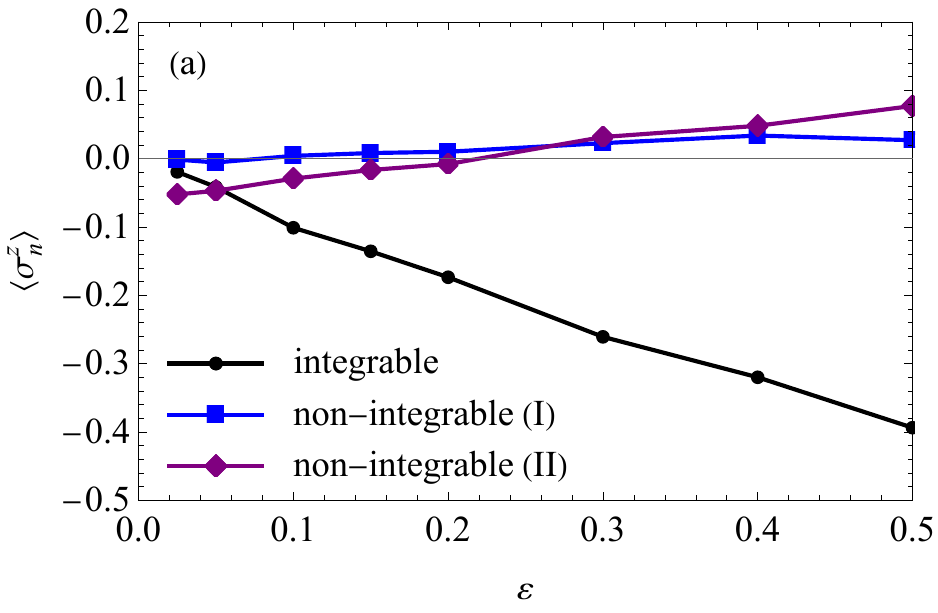}
    \includegraphics[width=7cm]{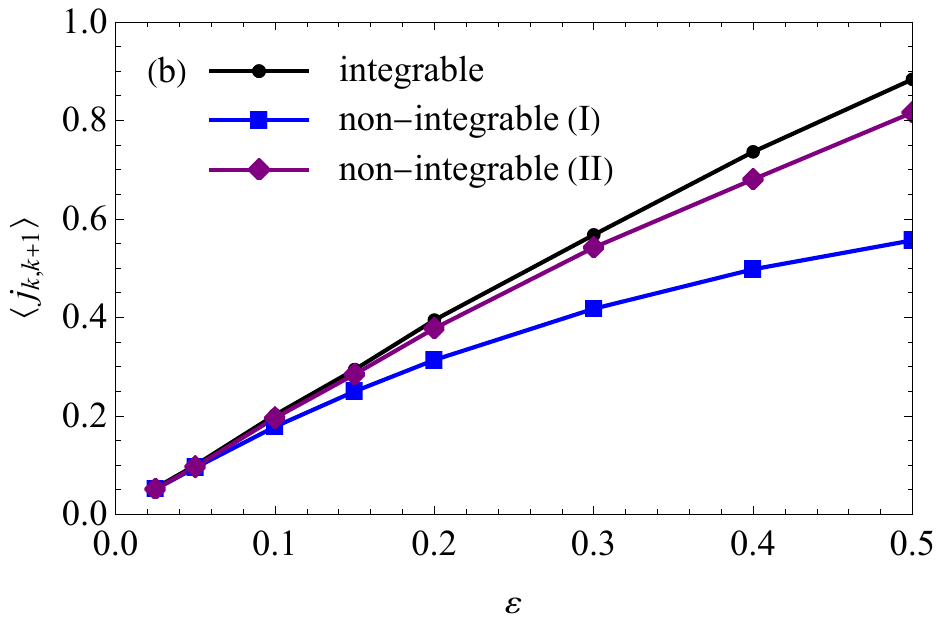}
    \caption{(a) The local magnetization at the impurity site $\langle\sigma_n^z\rangle$ and (b) the spin current $\langle j_{k,k+1}\rangle$ at $t=80$ as a function of the dissipation rate $\varepsilon$ for the integrable model and non-integrable models I and II ($N=12, n=7, \Delta=0.5, \xi=1, \varepsilon_{\rm R}=\varepsilon_{\rm L}=\varepsilon$) obtained by the quantum trajectory method. The initial state is the N\'{e}el state. We take the average over 4000 trajectories.}
    \label{fig:S^z epsilon}
\end{figure}

In Fig.~\ref{fig:S^z xi}, we compare the $\xi$ (impurity strength) dependence of the local magnetization at the impurity site $\langle\sigma_n^z\rangle$ (Fig.~\ref{fig:S^z xi}(a))
and the spin current $\langle j_{k,k+1}\rangle$ (Fig.~\ref{fig:S^z xi}(b)) between the integrable dissipative $XXZ$ model and non-integrable models I and II with $\Delta=0.5$. Note that $\langle j_{k,k+1}\rangle$ does not depend on $k$ (the site index) since the current is conserved and becomes spatially homogeneous in the steady state. We also show the exact solutions (Eqs.~(\ref{eq: S_k^z}) and (\ref{eq:current})) for the steady state of the integrable model by cross marks in Fig.~\ref{fig:S^z xi}, which agree well with the numerical results. As one can see, the integrable model has the largest modulus of the magnetization growing monotonically as a function of $\xi$, while the non-integrable models have relatively small magnetization. In particular, in the non-integrable model I the magnetization depends on $\xi$ non-monotonically. These results suggest that the integrability helps to induce local magnetization at the impurity site efficiently. To put it another way, the magnetization is suppressed in the non-integrable models due to scattering and heating. In the model II, the sign of the magnetization is opposite to the direction in which the energy decreases by the application of the magnetic field. This may be due to the effect of the large dissipation rate (see Fig.~\ref{fig:S^z epsilon}(a)), for which the current flows with a large amplitude and high-energy states are occupied. The spin current, on the other hand, remains almost unchanged against $\xi$ in the integrable model, while it generally varies when one breaks integrability (Fig.~\ref{fig:S^z xi}(b)). In the thermodynamics limit, the current becomes completely independent of $\xi$, as discussed in the previous section. This clearly demonstrates the role of integrability in the transport problem: The current is not hampered by the presence of impurities no matter how large the dissipation rate is as long as the integrability is maintained. It reminds us of the soliton motion in integrable systems, where two solitons keeps their shapes even after they collide with each other.

In Fig.~\ref{fig:S^z epsilon}, we compare the $\varepsilon$ dependence of the magnetization at the impurity site $\langle\sigma_n^z\rangle$
(Fig.~\ref{fig:S^z epsilon}(a)) and the spin current $\langle j_{k,k+1}\rangle$ (Fig.~\ref{fig:S^z epsilon}(b)) between the integrable model and the non-integrable model I and II with $\Delta=0.5$. In the integrable case, the magnetization is amplified as one increases the dissipation rate $\varepsilon$, whereas it is generally suppressed in the non-integrable models. The latter behavior is well expected since the magnetization is reduced by heating effects. In the non-integrable model I, the magnetization is accidentally small near $\varepsilon=0$. This may be due to the choice of the parameter $\xi=1$. As we vary the value of $\xi$, the magnetization grows to larger values (see Fig.~\ref{fig:S^z xi}(a)). In the non-integrable model II, the local magnetization appears already near $\varepsilon=0$ since the magnetic field is applied to the site $k=n$. This $\varepsilon$ dependence suggests a sharp difference between the current-induced magnetization and the trivial magnetic-field-induced magnetization. In the model II,
the local magnetization changes its sign at certain $\varepsilon$ instead of approaching zero in the large $\varepsilon$ limit. We have not fully understood the origin of this behavior, but it may be due to the effect of large amplitude of current that makes spins to occupy high-energy states. The current shows similar linear slopes among the three models in the weak $\varepsilon$ regime (Fig.~\ref{fig:S^z epsilon}(b)). As one increases $\varepsilon$, the current deviates from the linear dependence, and gradually shows nonlinear characteristics with smaller gradients.

\section{Concluding remarks}
In this paper, we have constructed the exact steady state of the Lindblad quantum master equation for the impurity-doped $XXZ$ spin model coupled to the dissipators at both ends in the matrix product form. 
Construction of the matrix product steady state is based on the bulk integrability, guaranteed by the existence of the $R$-matrix and the Lax operator which satisfies the fundamental commutation relation (FCR). 

We have calculated the expectation values for the magnetization profiles and the spin current from the exactly derived steady state in the matrix product form. We have found that local magnetization emerges at the position of the impurity when the current flows, which is in stark contrast to the usual situation that current generally suppresses magnetization via heating. This current-induced magnetization is shown to survive in the thermodynamic limit.
We have also proved that the spin current is independent of the impurity strength in the thermodynamic limit. This clearly
demonstrates the role of integrability in the transport problem. The current is not disturbed by
impurities no matter how strongly the system is driven by boundary dissipation as long as the bulk integrability is maintained. 
As the integrability is broken, the magnetization and current tend to be suppressed, which has been confirmed numerically by the quantum trajectory method.

Integrability-based construction of the steady state strongly motivates us to further study transport properties in other types of dissipative quantum models with bulk integrability. For example, one can introduce inhomogeneous parameters not only on the single site but on multiple sites without breaking integrability~\cite{bib:FZ97, bib:S94, bib:ZS97}. One can even randomly choose the parameters, which may implement integrable randomly disordered systems. Exact analysis of the quantum master equation on those models will help to understand transport properties and the role of integrability in interacting dissipative many-body systems with a random disorder.


\section*{Appendices}
\appendix

\section{Spin current around the impurity} \label{sec:spin_currents}
The spin current satisfies the continuity equation 
\begin{equation} \label{eq:continuity}
    \frac{d}{dt} \sigma_k^z = i[H,\,\sigma_k^z]
    = j_{k-1,k} - j_{k,k+1}
\end{equation}
at any site $k$. When the impurity locates at the $n$th site, we have the spin current 
\begin{equation}
    j_{k,k+1} = 4i (\sigma_k^+ \sigma_{k+1}^- - \sigma_k^- \sigma_{k+1}^+)
\end{equation}
for $k \neq n-1,n$. 
On the other hand, at the neighborhoods of the impurity we have
\begin{align}
    	&j_{n-1,n} = 4i \frac{1 - \Delta^2}{1 - \Delta^2 + \sinh^2\xi} \cosh\xi (\sigma_{n-1}^+ \sigma_{n}^- - \sigma_{n-1}^- \sigma_{n}^+) \nonumber \\
		&\hspace{10mm}+ 4 \frac{1}{1 - \Delta^2 + \sinh^2\xi} \Big( i \Delta \sinh^2\xi (\sigma_{n-1}^+ \sigma_{n+1}^- - \sigma_{n-1}^- \sigma_{n+1}^+) \nonumber \\
		&\hspace{44mm}- \Delta \sqrt{1 - \Delta^2} \sinh\xi (\sigma_{n-1}^+ \sigma_n^- + \sigma_{n-1}^- \sigma_n^+) \sigma_{n+1}^z \nonumber \\
		&\hspace{44mm}+ \sqrt{1 - \Delta^2} \sinh\xi \cosh\xi (\sigma_{n-1}^+ \sigma_n^z \sigma_{n+1}^- + \sigma_{n-1}^- \sigma_n^z \sigma_{n+1}^+) \Big),\\
	&j_{n,n+1} = 4i \frac{1 - \Delta^2}{1 - \Delta^2 + \sinh^2\xi} \cosh\xi (\sigma_{n}^+ \sigma_{n+1}^- - \sigma_{n}^- \sigma_{n+1}^+) \nonumber \\
		&\hspace{10mm}+ 4\frac{1}{1 - \Delta^2 + \sinh^2\xi} \Big( i \Delta \sinh^2\xi (\sigma_{n-1}^+ \sigma_{n+1}^- - \sigma_{n-1}^- \sigma_{n+1}^+) \nonumber \\
		&\hspace{44mm}- \Delta \sqrt{1 - \Delta^2} \sinh\xi \cdot \sigma_{n-1}^z (\sigma_{n}^+ \sigma_{n+1}^- + \sigma_{n}^- \sigma_{n+1}^+) \nonumber \\
		&\hspace{44mm}+ \sqrt{1 - \Delta^2} \sinh\xi \cosh\xi (\sigma_{n-1}^+ \sigma_n^z \sigma_{n+1}^- + \sigma_{n-1}^- \sigma_n^z \sigma_{n+1}^+) \Big)
\end{align}
for $\Delta\le 1$ from the continuity equation \eqref{eq:continuity}.

\section{Matrix product forms of $\Lambda$-local operators} \label{sec:MPforms}
Here we give the matrix product forms for expectation values of $\Lambda (\in \mathbb{N})$-local operators. We define $\Lambda$-local operators by the operator which non-trivially acts on $\Lambda$ consecutive sites. Noting that the Pauli matrices together with the identity operators form the basis of two-by-two matrices, any $\Lambda$-local operator is written as 
\begin{align}
    &O_{m+1,\dots,m+\Lambda} = \sum_{\alpha_j \in \{x,y,z,0\}} 
    c_{\alpha_{m+1},\dots,\alpha_{m+\Lambda}} 
    \sigma_{m+1}^{\alpha_{m+1}} \cdots \sigma_{m+\Lambda}^{\alpha_{m+\Lambda}},  
    \quad c_{\alpha_{m+1},\dots,\alpha_{,+\Lambda}} \in \mathbb{C}.
\end{align}
Since each Pauli matrix is the traceless operator, while the square of them coincides with the identity operator, we have 
\begin{align}
    &\langle O_{m+1,\dots,m+\Lambda} \rangle \nonumber 
    \\
    &= \sum_{\alpha_j \in \{x,y,z,0\}}
    c_{\alpha_{m+1},\dots,\alpha_{m+\Lambda}} 
    \frac{{_a}\langle 0| \otimes {_b}\langle 0| \prod_{x=1}^m \mathbb{L}_{a,b}^{0}(\lambda^s_x) \mathbb{L}_{a,b}^{\alpha_{m+1}}(\lambda^s_{m+1}) \cdots \mathbb{L}_{a,b}^{\alpha_{m+\Lambda}}(\lambda^s_{m+\Lambda}) \prod_{x=m+\Lambda+1}^N \mathbb{L}_{a,b}^0(\lambda^s_x) |0 \rangle_a \otimes |0 \rangle_b}
 {{_a}\langle 0| \otimes {_b}\langle 0| \prod_{x=1}^N \mathbb{L}_{a,b}^0(\lambda^s_x) |0 \rangle_a \otimes |0 \rangle_b}, 
\end{align}
where the set of spectral parameters $\{\lambda^s_x\}$ are the steady state spectral parameters defined in Section 3.

\section{Symmetrization of $\bm{L}^0$} \label{sec:symmetrization}
The existence of the basis which makes $\bm{L}_{a,b}^0(\lambda)$ symmetric and $\bm{L}_{a,b}^z(\lambda)$ anti-symmetric for $\lambda = i\pi/2$ is already verified in \cite{bib:P15}. By scaling the vectors as $|m,m \rangle \to \lambda_m|m,m \rangle$ in such a way that satisfies
\begin{equation}
	\frac{\lambda_{m+1}}{\lambda_m} |i\sin(\gamma(2s-m))|^2 = \frac{\lambda_m}{\lambda_{m+1}} |i\sin(\gamma(m+1))|^2, 
\end{equation}
it is straightforwardly obtained that $\bm{L}_{a,b}^0(\tfrac{i\pi}{2})$ is symmetric. Noting that $\bm{L}_{a,b}^z(\lambda)$ takes the form of 
\begin{align}
	\bm{L}_{a,b}^z(\lambda) 
	&= \sinh\overline{\lambda}\, \cosh\lambda \sum_{m=0}^{l-1} \cos(\gamma(\overline{s}-m)) \sin(\gamma(s-m)) |m,m \rangle \langle m,m| \nonumber\\
	&+ \cosh\overline{\lambda}\, \sinh\lambda \sum_{m=0}^{l-1} \sin(\gamma(\overline{s}-m)) \cos(\gamma(s-m)) |m,m \rangle \langle m,m| \nonumber\\
	&+ \frac{1}{2} \sum_{m=0}^{l-2} |\sin(\gamma(m+1))|^2\, |m,m \rangle \langle m+1,m+1| \nonumber\\
	&- \frac{1}{2} \sum_{m=0}^{l-2} |\sin(\gamma(2s-m))|^2\, |m+1,m+1 \rangle \langle m,m|, 
\end{align}
one obtains that all the diagonal elements vanishes at $\lambda = i\pi/2$. Thus $\bm{L}_{a,b}^z(\tfrac{i\pi}{2})$ is antisymmetric in the scaled basis. Note that $\bm{L}_{a,b}^z(\frac{i\pi}{2} + \xi)$ ($\xi \neq 0$) is not antisymmetric in the same basis, which produces the first term in \eqref{eq:asym_Sz}. 

Since $\bm{L}_{a,b}^0(\lambda + \xi)$ contains the parameter $\xi$ only in the diagonal terms, the same scaled basis makes $\bm{L}_{a,b}^0(\tfrac{i\pi}{2} + \xi)$ symmetric. It is easy to check that $[\bm{L}_{a,b}^0(\frac{i\pi}{2}),\,\bm{L}_{a,b}^0(\frac{i\pi}{2} + \xi)] = 0$, that is, $\bm{L}_{a,b}^0(\frac{i\pi}{2} + \xi)$ is simultaneously diagonalized with $\bm{L}_{a,b}^0(\frac{i\pi}{2})$.


\section*{Acknowledgements}
C. M. is supported by JSPS KAKENHI Grant Number JP18K13465 and Grant Number JP23K03244. 
N. T. is supported by JST FORESTO (Grant No.~JPMJFR2131) and KAKENHI (Grant No.~JP20K03811).

\bibliographystyle{abbrv}
\bibliography{references}

\end{document}